# AI Transparency in Academic Search Systems: An Initial Exploration

**Liu, Yifan** — School of Information, University of British Columbia, Canada | yiifan@student.ubc.ca
**Sullivan, Peter** — School of Information, University of British Columbia, Canada | prsull@student.ubc.ca
**Sinnamon, Luanne** — School of Information, University of British Columbia, Canada | luanne.sinnamon@ubc.ca

## ABSTRACT
As AI-enhanced academic search systems become increasingly popular among researchers, investigating their AI transparency is crucial to ensure trust in the search outcomes, as well as the reliability and integrity of scholarly work. This study employs a qualitative content analysis approach to examine the websites of a sample of 10 AI-enhanced academic search systems identified through university library guides. The assessed level of transparency varies across these systems: five provide detailed information about their mechanisms, three offer partial information, and two provide little to no information. These findings indicate that the academic community is recommending and using tools with opaque functionalities, raising concerns about research integrity, including issues of reproducibility and researcher responsibility.

## KEYWORDS
Academic Search System, AI Transparency, Artificial Intelligence, Academic Libraries

## INTRODUCTION
The rapid development of artificial intelligence (AI) is transforming work paradigms across society. This advancement creates opportunities and poses risks, which make transparency essential for ensuring trustworthy and comprehensible outcomes (Andrada et al., 2023; Larsson & Heintz, 2020). The Association for Computing Machinery (ACM) describes AI transparency as clear documentation of "the way in which specific datasets, variables, and models were selected for development, training, validation, and testing, as well as the specific measures that were used to guarantee data and output quality" (ACM, 2022, p. 3). Within the research and scholarly communication fields, researchers have long-established norms and means of sharing information about research methods and tools (Huvila & Sinnamon, 2022). However, the same norms for transparency in AI research tools are lacking (Lund et al., 2023). Definitions and standards for AI transparency vary widely and are not yet well established within academic and research contexts (Felzmann et al., 2020).

Academic search systems are algorithmic and/or AI-driven systems designed to facilitate the discovery, retrieval and use of research publications (Gusenbauer & Haddaway, 2021; Ortega, 2014). As academic search systems increasingly integrate AI to augment retrieval and synthesis functionalities (Heidt, 2023; Raymond, 2019), questions have been raised regarding how this AI integration impacts research (Lund et al., 2023). Recent studies mainly focus on evaluating the function, structure, coverage, and design of academic search systems (Ortega, 2014; Shafiq & Wani, 2018), as well as their role in scholarly practices, such as reducing information overload (Raymond, 2019). Concerns have been raised regarding bias, irreproducibility, hallucinations, opaque mechanisms, and inaccurately described limitations of LLM-assisted search systems (Gusenbauer, 2023; von Hippel & Buck, 2023). In the case of systematic reviews and meta-analyses, some search systems (e.g. Google Scholar) may be inappropriate (Gusenbauer, 2023). However, the needs and expectations with respect to AI transparency in academic search systems have not been thoroughly explored to date. This study takes some initial steps in this direction, assessing the extent to which widely used and recommended AI-enhanced academic search systems are transparent by examining publicly available information on the systems' mechanisms. We were guided by the commonsense research question: What would the average researcher using one of these systems be able to learn about how it works from readily available information?

## METHODS
To identify widely used systems for this analysis, we drew upon the recommendations of academic librarians. We searched Google, Bing, and DuckDuckGo using the query terms *library guide* and *AI search system*. Comparing the top 20 results from each search engine, we identified eight library guides that appeared in all three sets of results, namely from Georgetown University Library, University of Cambridge Library, University of Michigan Library, University of South Florida Libraries, Rutgers University Libraries, Carleton University MacOdrum Library, Oklahoma State University Libraries, and Texas A&M University Libraries. These libraries are geographically distributed, and the guides vary in scope, reflecting a diverse range of available AI-powered search tools. Based on the criteria of being an AI-enhanced academic search system, and appearing in at least three library guides, 10 systems out of 29 were selected for further analysis (Table 1). We consider general-purpose LLM-based chatbots including, ChatGPT, Perplexity, and Gemini to be out of scope based on our definition of academic search systems.

We employed a qualitative content analysis approach to examine the websites of the 10 systems. Existing studies on AI transparency distinguish between the nature of the algorithms and the data used in the training regimen (Andrada et al., 2023). We identified publicly available information regarding 1) the mechanisms of these systems, specifically how AI integration affects search results; 2) any described data used in the training regimen; and 3) details about the source databases.

**RESULTS**

In examining the system websites, we identified three kinds of information typically provided: the AI methods employed, the number of items indexed, and the source databases used (Table 1). The level of transparency and detail varies significantly across systems. Systems 1, 2, 3, and 4 provide substantial information on the AI methods and datasets, while systems 7, 8, and 9 offer little to no information. Systems 6 and 8 outline complex pipelines of components, posing additional challenges for transparency due to the lack of clear indication of component weighting and the absence of demonstrated efficiency for each component. A few systems (1, 5) include references to publications that describe the scientific process used to develop and evaluate system components (Cachola et al., 2020; Nicholson et al., 2021); however, these are not easily findable and may be outdated or only partially reflective of the current systems in use. While many systems list their capabilities, few provide information regarding their limitations. One exception is Elicit, which discusses its limitations, noting that as an early-stage tool, it is not 100% accurate and may miss nuances or misunderstand details. Additionally, it highlights general limitations of academic search tools, such as challenges in evaluating paper quality, the risk of confirmation bias, and variation in performance across different fields and research methods (*Elicit's Limitations*, n.d.).

| System | Methods Employed | Works Indexed | Database Used |
| --- | --- | --- | --- |
| (1) Semantic Scholar | LLM | 214 M | Semantic Scholar Paper Corpus |
| (2) Connected Papers | Force Directed Graph | Total Not Provided | Semantic Scholar Paper Corpus |
| (3) Consensus | LLM | 200 M | Semantic Scholar Paper Corpus |
| (4) Elicit | LLM | 126 M | Semantic Scholar Paper Corpus |
| (5) Scite.ai | LLM | 187 M | A Licensed and Open Access Dataset |
| (6) Keenious | Ensemble Model | 100 M | Not Described |
| (7) SciSpace | Not Described | Not Described | Web of Science Core Collection Database |
| (8) Paper Digest | Optional LLM, Graph and Deep Learning | Not Described | Not Described |
| (9) Scholarcy | Not Described | Not Described | Not Described |
| (10) Research Rabbit | Not Described | Not Described | Not Described |

Table 1. Overview of Transparency Features in AI-Enhanced Academic Search Systems

**DISCUSSION AND CONCLUSION**

The average user would be hard-pressed to understand or explain the outcomes of many AI-enhanced academic search systems due to insufficient information or a mismatch between the information provided and the users' expertise. While these systems promise significant advancements in research support, their lack of transparency poses a threat to research integrity and reliability, highlighting the need for a more comprehensive examination to develop standardized guidelines for AI transparency in academic search systems. At a minimum, these guidelines should provide clear and substantial documentation of AI mechanisms, identify the database/s used, and the works indexed. Search system developers should clarify the functions and methodology and ensure that scientific works describing their systems are directly linked. Further, academic librarians recommending these systems should consider systems' AI transparency and provide patrons with the opportunity to understand how these systems function. AI literacy and transparency expectations should be included when educating patrons on the use of academic search tools. Researchers bear the responsibility of employing clearly defined literature search methods, as reliance on opaque search systems can exacerbate existing biases. While we recognize that this is a rapidly developing area and this study is not comprehensive, it represents an initial step towards further exploration on the nature and impacts of AI-enhanced academic search systems.

**GENERATIVE AI USE**

AI-enhanced academic search systems were the subject of this study. We did not use generative AI tools/services in any other steps of analysis or writing. The authors assume all responsibility for the content of this submission.

**AUTHOR ATTRIBUTION**

Yifan Liu: conceptualization, data curation, formal analysis, methodology, project administration, validation, writing – original draft, writing – review and editing; Peter Sullivan: formal analysis, validation, writing – review and editing; Luanne Sinnamon: conceptualization, supervision, writing – review and editing

**REFERENCES**

ACM Technology Policy Council (2022) Statement on principles for responsible algorithmic systems. https://www.acm.org/binaries/content/assets/public-policy/final-joint-ai-statement-update.pdf

Andrada, G., Clowes, R. W., & Smart, P. R. (2023). Varieties of transparency: Exploring agency within AI systems. *AI & SOCIETY*, *38*(4), 1321–1331. https://doi.org/10.1007/s00146-021-01326-6

Cachola, I., Lo, K., Cohan, A., & Weld, D. (2020). TLDR: Extreme summarization of scientific documents. In T. Cohn, Y. He, & Y. Liu (Eds.), *Findings of the Association for Computational Linguistics: EMNLP 2020* (pp. 4766–4777). Association for Computational Linguistics. https://doi.org/10.18653/v1/2020.findings-emnlp.428

*Elicit's Limitations*. (n.d.). Retrieved May 18, 2024, from https://support.elicit.com/en/articles/549569

Felzmann, H., Fosch-Villaronga, E., Lutz, C., & Tamò-Larrieux, A. (2020). Towards transparency by design for artificial intelligence. *Science and Engineering Ethics*, *26*(6), 3333–3361. https://doi.org/10.1007/s11948-020-00276-4

Gusenbauer, M. (2023). Audit AI search tools now, before they skew research. *Nature*, *617*(7961), 439–439. https://doi.org/10.1038/d41586-023-01613-w

Gusenbauer, M., & Haddaway, N. R. (2021). What every researcher should know about searching – clarified concepts, search advice, and an agenda to improve finding in academia. *Research Synthesis Methods*, *12*(2), 136–147. https://doi.org/10.1002/jrsm.1457

Huvila, I., & Sinnamon, L. (2022). Sharing research design, methods and process information in and out of academia. *Proceedings of the Association for Information Science and Technology*, *59*(1), 132–144. https://doi.org/10.1002/pra2.611

Larsson, S., & Heintz, F. (2020). Transparency in artificial intelligence. *Internet Policy Review*, *9*(2). https://policyreview.info/concepts/transparency-artificial-intelligence

Lund, B. D., Wang, T., Mannuru, N. R., Nie, B., Shimray, S., & Wang, Z. (2023). ChatGPT and a new academic reality: Artificial intelligence-written research papers and the ethics of the large language models in scholarly publishing. *Journal of the Association for Information Science and Technology*, *74*(5), 570–581. https://doi.org/10.1002/asi.24750

Nicholson, J. M., Mordaunt, M., Lopez, P., Uppala, A., Rosati, D., Rodrigues, N. P., Grabitz, P., & Rife, S. C. (2021). scite: A smart citation index that displays the context of citations and classifies their intent using deep learning. *Quantitative Science Studies*, *2*(3), 882–898. https://doi.org/10.1162/qss_a_00146

Ortega, J. L. (2014). *Academic search engines: A quantitative outlook*. Chandos Publishing. https://doi.org/10.1533/9781780634722.1

Raymond, D. (2019). Using artificial intelligence to combat information overload in research. *IEEE Pulse*, *10*(1), 18–21. https://doi.org/10.1109/MPULS.2018.2885843

Shafiq, H., & Wani, Z. A. (2018). Assessment of search interface of information retrieval systems A case study of select academic databases. *2018 5th International Symposium on Emerging Trends and Technologies in Libraries and Information Services (ETTLIS)*, 45–53. https://doi.org/10.1109/ETTLIS.2018.8485262

von Hippel, P. T., & Buck, S. (2023). Improve academic search engines to reduce scholars' biases. *Nature Human Behaviour*, *7*(2), Article 2. https://doi.org/10.1038/s41562-022-01518-0